\documentclass[aps,pre,twocolumn,showpacs]{revtex4-1}

\usepackage{graphicx}
\usepackage{color}

\begin{document}

\title{Flux-based classification of reactions reveals a functional bow-tie
organization of complex metabolic networks}

\author{Shalini Singh$^{1,2}$}
\author{Areejit Samal$^{1,3,4}$}
\author{Varun Giri$^{1}$}
\author{Sandeep Krishna$^{5}$}
\author{Nandula Raghuram$^{6}$}
\author{Sanjay Jain$^{1,7,8}$}
\email{jain@physics.du.ac.in}

\affiliation{$^1$Department of Physics and Astrophysics, University
of Delhi, Delhi 110007, India} \affiliation{$^2$ Department of
Genetics, University of Delhi, South Campus, New Delhi, India}
\affiliation{$^3$Max Planck Institute for Mathematics in the
Sciences, Inselstrasse 22, D-04103 Leipzig, Germany}
\affiliation{$^4$Laboratoire de Physique Th\'{e}orique et
Mod\`{e}les Statistiques, CNRS and Univ Paris-Sud, UMR 8626, F-91405
Orsay, France} \affiliation{$^5$National Centre for Biological
Sciences, UAS-GKVK Campus, Bangalore 560065, India}
\affiliation{$^6$School of Biotechnology, GGS Indraprastha
University, Dwarka, New Delhi 110078, India} \affiliation{$^7$
Jawaharlal Nehru Centre for Advanced Scientific Research, Bangalore
560064, India} \affiliation{$^8$Santa Fe Institute, 1399 Hyde Park
Road, Santa Fe, NM 87501, USA}


\begin{abstract}
Unraveling the structure of complex biological networks and relating
it to their functional role is an important task in systems biology.
Here we attempt to characterize the functional organization of the
large-scale metabolic networks of three microorganisms. We apply
flux balance analysis to study the optimal growth states of these
organisms in different environments. By investigating the
differential usage of reactions across flux patterns for different
environments, we observe a striking bimodal distribution in the
activity of reactions. Motivated by this, we propose a simple
algorithm to decompose the metabolic network into three
sub-networks. It turns out that our reaction classifier which is
blind to the biochemical role of pathways leads to three
functionally relevant sub-networks that correspond to input, output
and intermediate parts of the metabolic network with distinct
structural characteristics. Our decomposition method unveils a
functional bow-tie organization of metabolic networks that is
different from the bow-tie structure determined by graph-theoretic
methods that do not incorporate functionality.
\end{abstract}

\maketitle

\section{Introduction}
Biological systems provide many examples of the intricate
relationship between the structure and functionality of complex
networks
\cite{hartwell1999molecular,bornholdt2003handbook,barabasi2004network,wagner2005robustness,sneppen2005physics,alon2006introduction,kaneko2006life}.
Cellular metabolism is a complex biochemical network of several
hundred metabolites that are processed and interconverted by
enzyme-catalyzed reactions
\cite{heinrich1996regulation,jeong2000large,wagner2001small,ma2003connectivity,csete2004bow,palsson2006systems}.
Metabolic networks have a dynamic flexibility that enables organisms
to survive under diverse environmental conditions. A key goal of
systems biology is to unveil the functional organization of
metabolic networks explaining their system-level response to
different environments. To this end, we have attempted to decompose
metabolic networks into functionally relevant sub-networks. Flux
balance analysis (FBA) has been widely used to harness the knowledge
of large-scale metabolic networks and investigate genotype-phenotype
relationships
\cite{price2004genome,feist2008growing,oberhardt2009applications}.
FBA has been successful in predicting the growth and deletion
phenotypes of organisms
\cite{edwards2001silico,ibarra2002escherichia,segre2002analysis}.
Reaction fluxes carry information about the flows on metabolic
networks and, as such, describe the functional use of the network by
the organism. In this paper, we have used this information to
decompose the network into functionally relevant sub-networks.

The paper is organized as follows: In section II we describe the modelling framework in which we study metabolic networks. In section III we discuss the classification of active reactions in metabolic networks into three categories by an algorithm that is blind to their biochemical roles. Section IV shows that the three categories are functionally relevant for the organism. In section V we compare the bow-tie architecture obtained by our functional classification of reactions with that obtained by graph-theoretic methods that do not employ functional information. In the last section we conclude with a summary.


\begin{table*}
\caption{\label{stat} Comparison of the three metabolic networks:
\emph{E. coli}, \emph{S. cerevisiae} and \emph{S. aureus}.}
\begin{tabular}{|c|c|c|c|}
\hline Property &   \emph{E. coli} & \emph{S. cerevisiae} &  \emph{S. aureus}\\
\hline Number of metabolites & 761 & 1061 & 648\\
\hline Number of reactions in the model  &  931 & 1149 &  641\\
\hline Number of one-sided reactions in the equivalent network &
1167 & 1576 & 863\\
\hline Number of external metabolites & 143 & 116 & 84\\
\hline Number of organic external metabolites (carbon sources) & 131
& 107 & 68\\
\hline Number of biomass metabolites & 49 & 42 & 56\\
\hline Number of feasible minimal environments & 89 & 43 & 27\\
\hline Number of active reactions & 585 & 482 & 418\\
\hline Number of reactions in category I &  185 & 89 & 84\\
\hline Number of reactions in category IIa & 147 & 117 & 194\\
\hline Number of reactions in category IIb & 42 & 46 & 28\\
\hline Number of reactions in category III & 211 & 230 & 112\\
\hline
\end{tabular}
\end{table*}


\section{The modelling framework}

\subsection{Flux balance analysis (FBA)}
Flux balance analysis (FBA) is a computational approach widely used
to analyze the capabilities of genome-scale metabolic networks
\cite{price2004genome,feist2008growing,oberhardt2009applications}.
The stoichiometric matrix ${\bf S}$ encapsulates the stoichiometric
coefficients of different metabolites involved in various reactions
of the metabolic network. FBA primarily uses structural information
of the metabolic network contained in this matrix ${\bf S}$ to
predict the possible steady state flux distribution of all reactions
and the maximum growth rate of an organism. In any metabolic steady
state, the metabolites achieve a dynamic mass balance wherein the
vector ${\bf v}$ of fluxes through the reactions satisfies the
following equation representing the stoichiometric and mass balance
constraints:
\begin{equation}
\label{steadystate} {\bf S}.{\bf v} = 0 .
\end{equation}
Equation \ref{steadystate} is an under-determined linear system of
equations relating various reaction fluxes in genome-scale metabolic
networks leading to a large solution space of allowable fluxes. The
space of allowable solutions can be reduced by incorporating
thermodynamic and enzyme capacity constraints. To obtain a
particular solution, linear programming is used to find a set of
flux values - a particular flux vector ${\bf v}$ - that maximizes a
biologically relevant linear objective function $Z$. The linear
programming formulation of the FBA problem can be written as:
\begin{equation}
\label{fba}\mbox{max}\ Z\ =\ \mbox{max}\ \{ {\bf c^{T}}{\bf v}|{\bf
S}.{\bf v}=0, {\bf a}\le{\bf v}\le{\bf b}\} ,
\end{equation}
where vectors ${\bf a}$ and ${\bf b}$ contain the lower and upper
bounds of different fluxes in ${\bf v}$ and the vector ${\bf c}$
corresponds to the coefficients of the objective function $Z$. In
FBA, the objective function $Z$ is usually taken to be the growth
rate of the organism. The environment, or medium, is defined in this
approach by the components of ${\bf a}$ and ${\bf b}$ corresponding
to the transport reactions, which determine, in particular, the set
of metabolites whose uptake is allowed.

\subsection{Large-scale metabolic networks}
In this work, we have analyzed the large-scale metabolic networks of
three microorganisms: \emph{Escherichia coli} (version iJR904
\cite{reed2003expanded}), \emph{Saccharomyces cerevisiae} (version
iND750 \cite{duarte2004reconstruction}) and \emph{Staphylococcus
aureus} (version iSB619 \cite{becker2005genome}). Table \ref{stat}
gives the number of metabolites and reactions in the metabolic
networks of these three organisms. The metabolic networks contain
internal and transport reactions. Internal reactions occur within
the cell boundary. Transport reactions represent processes involving
import or export of metabolites across the cell boundary. Each model
also contains a pseudo biomass reaction that simulates the drain of
various biomass precursor metabolites for growth in the specific
organism. Starting from the published metabolic network, we obtain
an equivalent reaction network as follows: Every reversible reaction
in the network is converted into two one-sided (irreversible)
reactions so that all reaction fluxes in the equivalent system are
positive. A few reactions appear in duplicate in these networks, and
only a single copy of each reaction is kept in the equivalent
network. The equivalent metabolic network is a reaction set
consisting of $N$ unique one-sided reactions where $N$ is 1167, 1576
and 863 for \emph{E. coli}, \emph{S. cerevisiae} and \emph{S.
aureus}, respectively (cf. Table \ref{stat}).

\subsection{Feasible minimal environments and associated flux vectors}
In this work, we have considered minimal environments each
characterized by the presence of a limited amount of one organic
carbon source (external nutrient metabolite) along with unlimited
amounts of the following inorganic metabolites: ammonia, iron,
potassium, protons, pyrophosphate, sodium, sulfate, water and
oxygen. The number of environments we consider for each organism
thus coincides with the number of organic external metabolites
(carbon sources) in its metabolic network (cf. Table \ref{stat}). We
used FBA to determine the set of minimal environments supporting
growth in the metabolic networks of \emph{E. coli}, \emph{S.
cerevisiae} and \emph{S. aureus}. A minimal environment is termed as
\emph{feasible} if the growth rate predicted by FBA is nonzero for
that carbon source. The number $M$ of feasible minimal environments
in \emph{E. coli}, \emph{S. cerevisiae} and \emph{S. aureus} was
obtained to be 89, 43 and 27, respectively (cf. Table \ref{stat})
\cite{samal2006low}. For each organism, and for each feasible
minimal environment for that organism, we obtained an
$N$-dimensional optimal flux vector ${\bf v}$ maximizing growth in
the metabolic network of the organism, and whose component $v_j$
gives the flux of reaction $j$. For every organism, this led to a
set of $M$ flux vectors corresponding to $M$ feasible minimal
environments which were stored in the form of a matrix ${\bf
V}$$=$($v_j^{\alpha}$) of dimensions $N$$\times$$M$ where the rows
($j$=1,2,$\ldots$,$N$) correspond to different reactions in network
and columns ($\alpha$=1,2,$\ldots$,$M$) to different feasible
minimal environments. $v_j^{\alpha}$ is defined as the flux of
reaction $j$ in the optimal flux vector ${\bf v}$ obtained for
environment $\alpha$.

\subsection{Active reactions}
A given reaction $j$ is termed as \emph{active} in an environment
$\alpha$ if $v_j^{\alpha}$$>$$0$. The \emph{activity} $m$ of a
reaction denotes the number of minimal environments in which the
reaction is active. The activity $m$ for a reaction ranges from 0 to
$M$ with $M$ equal to 89, 43 and 27 for \emph{E. coli}, \emph{S.
cerevisiae} and \emph{S. aureus}, respectively. A reaction $j$ is
termed as active in an organism if $m$$\ge$1 (i.e., if it is active
in at least one feasible minimal environment for that organism). The
number of active reactions in \emph{E. coli}, \emph{S. cerevisiae}
and \emph{S. aureus} was obtained to be 585, 482 and 418,
respectively (cf. Table \ref{stat}). This paper primarily focuses on
decomposing this set of active reactions into functionally relevant
sub-networks.

\begin{figure*}
\includegraphics[width=12cm]{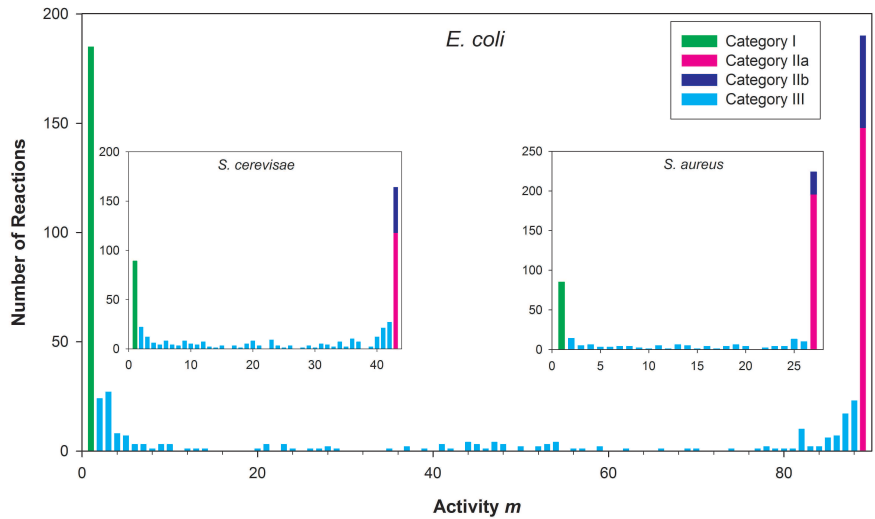}
\caption{\label{Fig1} {\bf The histogram of activity of reactions in
the \emph{E. coli} metabolic network.} The bars show the number of
reactions that have an activity $m$ where $m$ ranges from 1 to 89
feasible minimal environments in the \emph{E. coli} metabolic
network. The green bar represents 185 category I reactions which are
once-active. The pink bar represents 147 category IIa reactions (a
subset of 189 always active category II reactions) that have fluxes
perfectly correlated across environments. The deep blue bar
represents 47 category IIb reactions that account for the remaining
category II reactions. The light blue bars account for 211 category
III reactions with intermediate activity. {\bf Insets:} Histograms
of activity of reactions in \emph{S. cerevisiae} and \emph{S.
aureus}. The three categories of reactions in \emph{S. cerevisiae}
and \emph{S. aureus} were defined in a manner similar to \emph{E.
coli}.}
\end{figure*}

\begin{figure*}
\includegraphics[width=10cm]{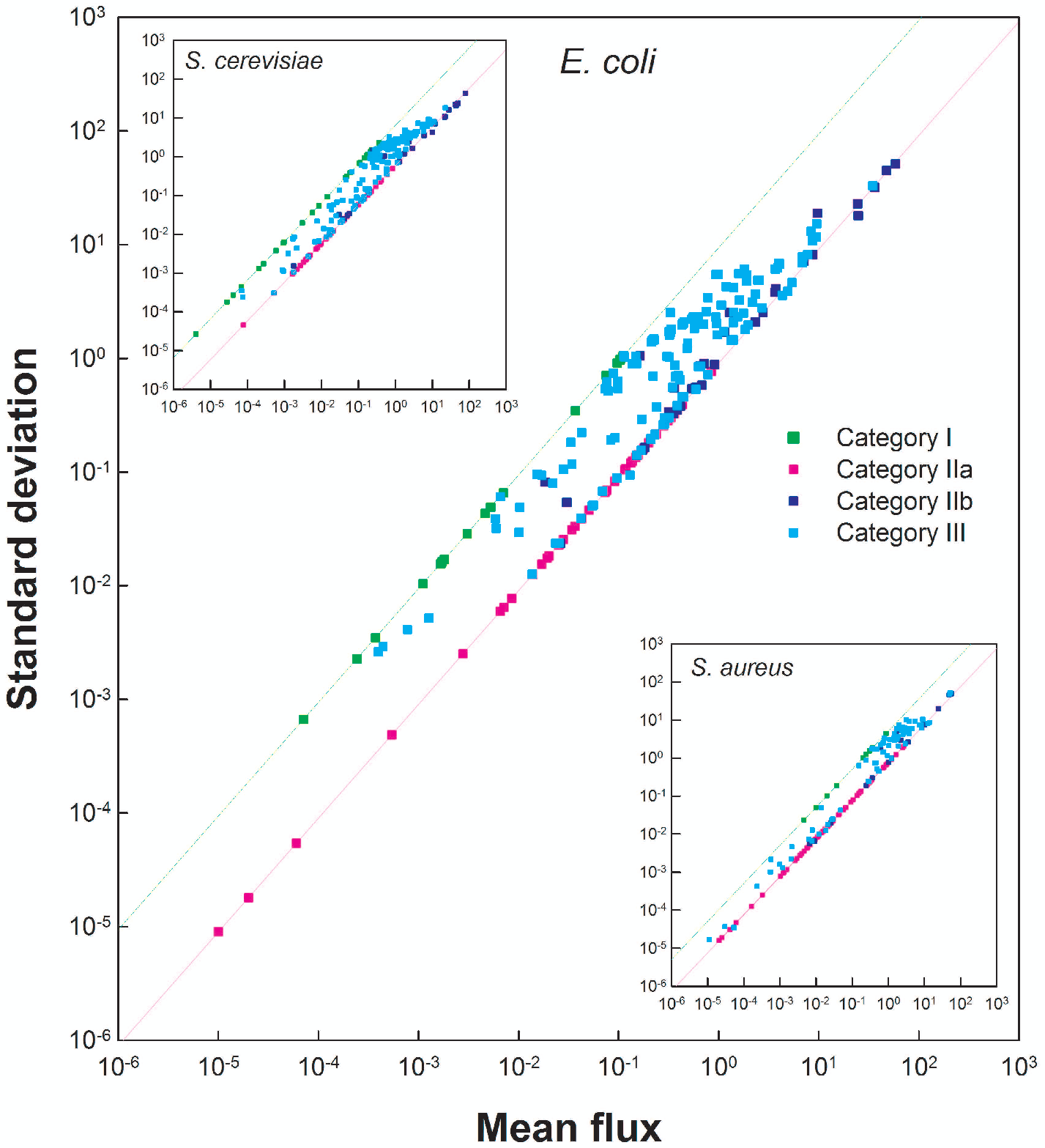}
\caption{\label{Fig2} {\bf Standard deviation versus mean flux of
active reactions in the \emph{E. coli} metabolic network.} The plot
shows standard deviation $\sigma$ versus mean flux $\langle v
\rangle$ of the 585 active reactions in \emph{E. coli} metabolic
network across $M=89$ feasible minimal environments on a logarithmic
scale. The green, pink, dark blue and cyan dots represent category
I, IIa, IIb and III reactions, respectively. The three categories of
reactions show up quite distinctly (upper line, category I; lower
line, category IIa; with category IIb and category III in between
the two lines). The upper line is the expected curve
$\sigma=(M-1)^{1/2}\langle v \rangle$ for category I reactions. The
lower line is the expected curve $\sigma=b\langle v \rangle$ for
perfectly correlated category IIa reactions with $b=0.98\pm0.1$
obtained via best fit to the data. {\bf Insets:} Scatter plots of
$\sigma$ versus $\langle v \rangle$  of active reactions in \emph{S.
cerevisiae} and \emph{S. aureus} metabolic networks.}
\end{figure*}

\section{Classification of active reactions}
We ask the question: How does the activity of a reaction vary across
different environments? To address this question, we determine the
frequency distribution of the activity of reactions in an organism.
Fig. 1 shows the histogram of the activity of reactions in the
\emph{E. coli} metabolic network. The distribution is bimodal. Most
reactions in \emph{E. coli} are either once-active ($m$=1) or always
active ($m$=89); the number of reactions for any given intermediate
activity $m$ in the range 1$<$$m$$<$89 is small. Thus, the largest
number of active reactions in the metabolic network are used in
either one environment or in all environments. The histograms of
activity of reactions in \emph{S. cerevisiae} and \emph{S. aureus}
also have a pattern similar to that in \emph{E. coli} (cf. Fig. 1).
The frequency distribution of activity of reactions in the three
organisms suggests a natural classification of active reactions into
three categories:
\begin{itemize}
\item[(a)] Category I reactions or once-active reactions ($m$=1)
\item[(b)] Category II reactions or always active reactions ($m$=$M$)
\item[(c)] Category III reactions with intermediate activity (1$<$$m$$<$$M$)
\end{itemize}

\subsection{Sub-classification based on correlation of reaction
fluxes} Clustering of gene expression data using the correlation
coefficient has been successful in predicting regulatory modules
associated with a biological function across diverse conditions
\cite{eisen1998cluster}. We used the correlation coefficient to
identify sets of reactions whose fluxes are correlated across
different environments. We used the set of $M$ flux vectors
corresponding to $M$ feasible minimal environments contained in
matrix ${\bf V}=(v_j^{\alpha})$ to obtain the matrix ${\bf C}= (C_{jk})$
where $C_{jk}$ is the correlation coefficient between two active
reactions $j$ and $k$ and is given by:
\begin{eqnarray}
\label{clustering} C_{jk} = \frac{1}{M}\sum_{\alpha=1}^M
\frac{v_j^\alpha v_k^\alpha}{\phi_j \phi_k}, \\
\mbox{where} \
\phi_j=\sqrt{\frac{1}{M}\sum_{\alpha=1}^M{v_j^\alpha}^2}. \nonumber
\end{eqnarray}

If $C_{jk}=1$ then reactions $j$ and $k$ are perfectly correlated
with each other in the given set of environments. Perfect clusters
in metabolic networks are maximal sets of reactions that are
perfectly correlated to each other pairwise. Perfect clusters are
similar to enzyme subsets
\cite{pfeiffer1999metatool,stelling2002metabolic}, correlated
reaction sets \cite{papin2002extreme,reed2004genome} or fully
coupled sets \cite{burgard2004flux} which have been used to detect
modules in metabolic networks.

We use Eq. \ref{clustering} to identify perfect clusters in
metabolic networks of \emph{E. coli}, \emph{S. cerevisiae} and
\emph{S. aureus}. In particular, a large perfect cluster of 147
reactions was found in \emph{E. coli} that is a subset of category
II reactions. We refer to this subset of perfectly correlated
reactions within category II as category IIa reactions. The
remaining 42 category II reactions that are always active but not
perfectly clustered with category IIa reactions are part of category
IIb. Similarly, large prefect clusters of sizes 117 and 194 were
found in category II reactions of \emph{S. cerevisiae} and \emph{S.
aureus}, respectively. In Fig. 1, category IIa and IIb reactions are
shown in pink and blue colours, respectively. We have shown
elsewhere that perfect clusters are metabolic modules that can be
explained by studying the connectivity of their constituent
metabolites \cite{samal2006low}.

Note that we have used a single optimal flux vector ${\bf v}$
obtained using FBA for each of the $M$ feasible minimal environments
to determine the activity of a reaction and the set of active
reactions in the metabolic network of an organism. However, it is
well known that there exist multiple flux vectors or alternate
optimal solutions in most large-scale metabolic networks that
maximize growth in a given environment
\cite{lee2000recursive,mahadevan2003effects,reed2004genome,samal2008conservation}.
In principle, due to the presence of alternate optima, the set of
active reactions can change depending on the choice of the flux
vectors. In Appendix A, we show the robustness of our reaction
categories to the presence of alternate optima.

\subsection{Scatter plot of standard deviation versus mean flux of
reactions across environments discriminates the three categories}
For each active reaction, following Almaas {\it et al}
\cite{almaas2004global}, we have calculated the mean flux $\langle v
\rangle$  and the standard deviation $\sigma$ around this mean by
averaging the flux of the reaction over $M$ feasible minimal
environments. Fig. 2 shows the scatter plot of  $\sigma$ versus
$\langle v \rangle$ for active reactions in \emph{E. coli}. From
Fig. 2, we can distinguish between categories I, II and III,
respectively, as they show up quite distinctly (upper line, category
I; lower line, category IIa; with category IIb and category III
largely in between the two lines). The upper line in Fig. 2 is the
expected curve $\sigma=(M-1)^{1/2}\langle v \rangle$ for category I
reactions and the lower line is the curve $\sigma=b\langle v
\rangle$, where $b$ is obtained via best fit of data for category
IIa reactions. Appendix B gives the derivation of the relation
between $\sigma$ and $\langle v \rangle$ for category I and IIa
reactions. Thus, we find that the three categories of reactions are
distinct from each other by virtue of their statistical properties.


\begin{figure*}
\includegraphics[width=16cm]{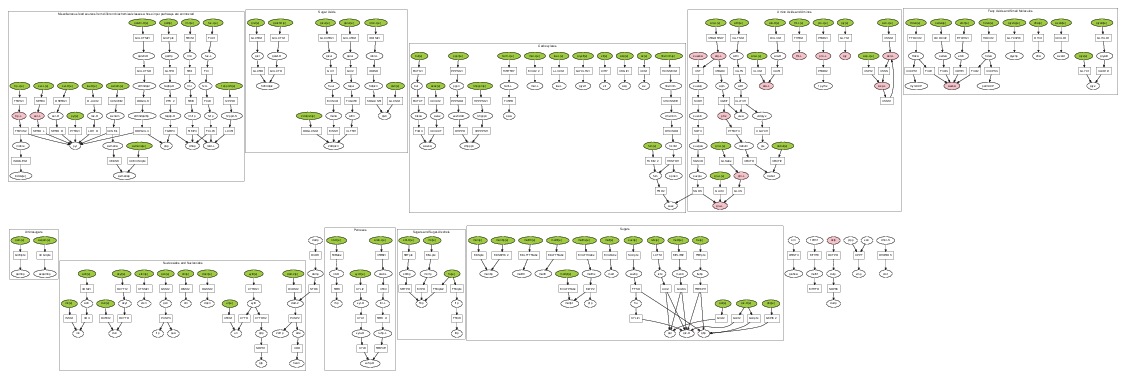}
\caption{\label{Fig3} {\bf Category I reactions in \emph{E. coli}.}
This figure shows the bipartite graph of 185 category I reactions in
\emph{E. coli}. Rectangles represent reactions and ovals
metabolites. External nutrient metabolites (organic carbon sources)
are depicted in green and biomass metabolites in pink. For
convenience, we have chosen to omit the high degree currency
metabolites (such as ATP) from the figure in order to reduce clutter
and focus on the biochemically relevant transformation in each
reaction. Abbreviation of metabolites and reactions are as in iJR904
model \cite{reed2003expanded}. The figure was drawn using Graphviz
software \cite{ellson2002graphviz}. The high resolution electronic
version of this figure can be zoomed in to read node labels and
biochemical categories of boxes. We have classified the external
metabolites and grouped together their input pathways in boxes based
on biochemical similarity. }
\end{figure*}

\begin{figure*}
\includegraphics[width=7cm]{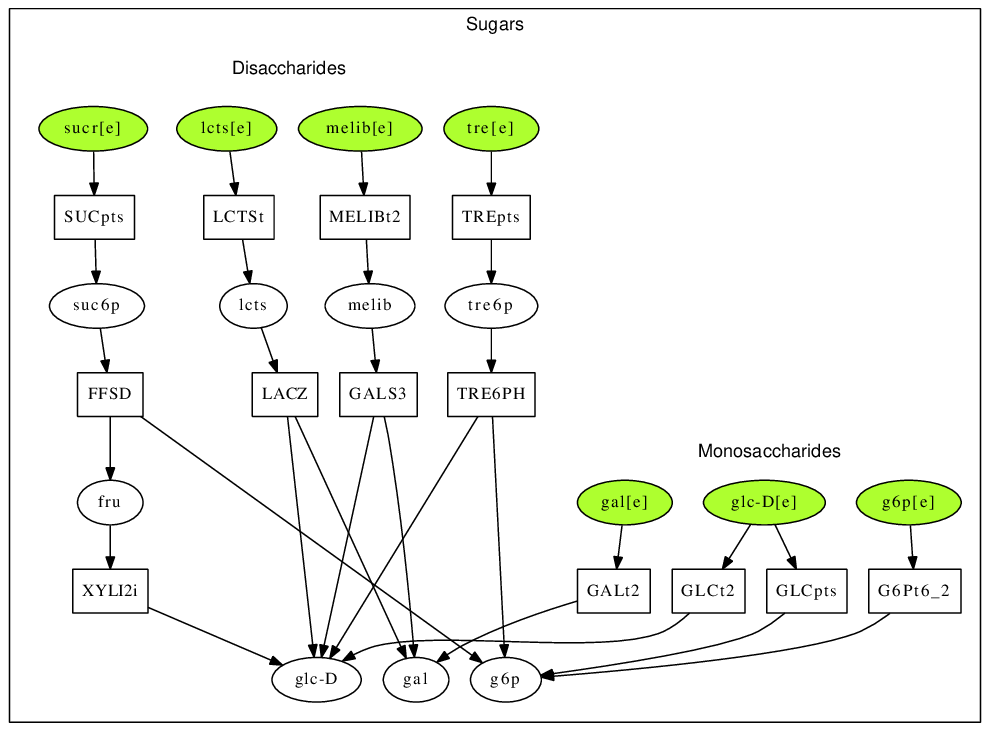}
\caption{\label{Fig4} {\bf A small portion of category I sub-network
in \emph{E. coli} showing sugar input pathways.} The figure shows
category I reactions in the input pathways for external nutrient
metabolites classified into the biochemical category `Sugars'. Two
kinds of sugars are shown here: monosaccharides and disaccharides.
The input pathways for 7 external sugar metabolites fan-in
downstream into 3 monosaccharide metabolites which occur at the
boundary between category I and III sub-networks. Conventions are
the same as in Figure 3.}
\end{figure*}


\section{Functional relevance of the three categories of reactions}
Until now our classification of active reactions into the three
categories was solely motivated by the activity of  reactions in
\emph{E. coli}, \emph{S. cerevisiae} and \emph{S. aureus} with two
very prominent peaks for once-active and always active reactions
(cf. Fig. 1). However, we now show that our three categories I, II,
and III obtained using a computational algorithm blind to the
biochemical nature of pathways corresponds to the input, output and
intermediate sub-networks, respectively. Thus, each category of
reactions is a sub-network with a distinct functional role in
metabolism.

\subsection{Category I: Fan-in of input pathways} Fig. 3 shows the
sub-network of all 185 category I reactions in \emph{E. coli}. The
figure shows a number of essentially linear paths of one to about
five reactions starting from an external nutrient metabolite, often
converging to some other metabolite. These are the input pathways of
those metabolites, typically starting from their transport reaction
that brings them into the cell, and subsequent catabolic reactions
that break them down into a smaller set of metabolites. Input
pathways of 86 out of the 89 external nutrient metabolites (carbon
sources) characterizing different feasible minimal environments are
contained in category I, thereby implying that category I
essentially covers all the input pathways of metabolism. Similarly,
we find that category I reactions in \emph{S. cerevisiae} and
\emph{S. aureus} contain input pathways for most external nutrient
metabolites characterizing different feasible minimal environments.
Thus, category I essentially corresponds to input part of the
metabolic network.

Fig. 4 shows a portion of category I reactions belonging to sugar
input pathways in \emph{E. coli} where several external sugar
metabolites converge downstream into a few intermediate metabolites.
Thus, the input pathways in category I exhibit the \emph{fan-in}
property whereby diverse external nutrient metabolites are first
catabolized into a smaller set of intermediate metabolites before
being drawn into the interior of the metabolic network. Usually the
external nutrients whose input pathways converge to a common
metabolite belong to the same biochemical class (cf. Figures 3 and
4). Fig. 3 contains a number of disconnected subgraphs each
describing the input pathways of one or more biochemically similar
metabolites; these disconnected paths get connected to the larger
metabolic network via further downstream reactions that belong to
other categories and are not shown in Fig. 3.

\begin{figure*}
\includegraphics[width=16cm]{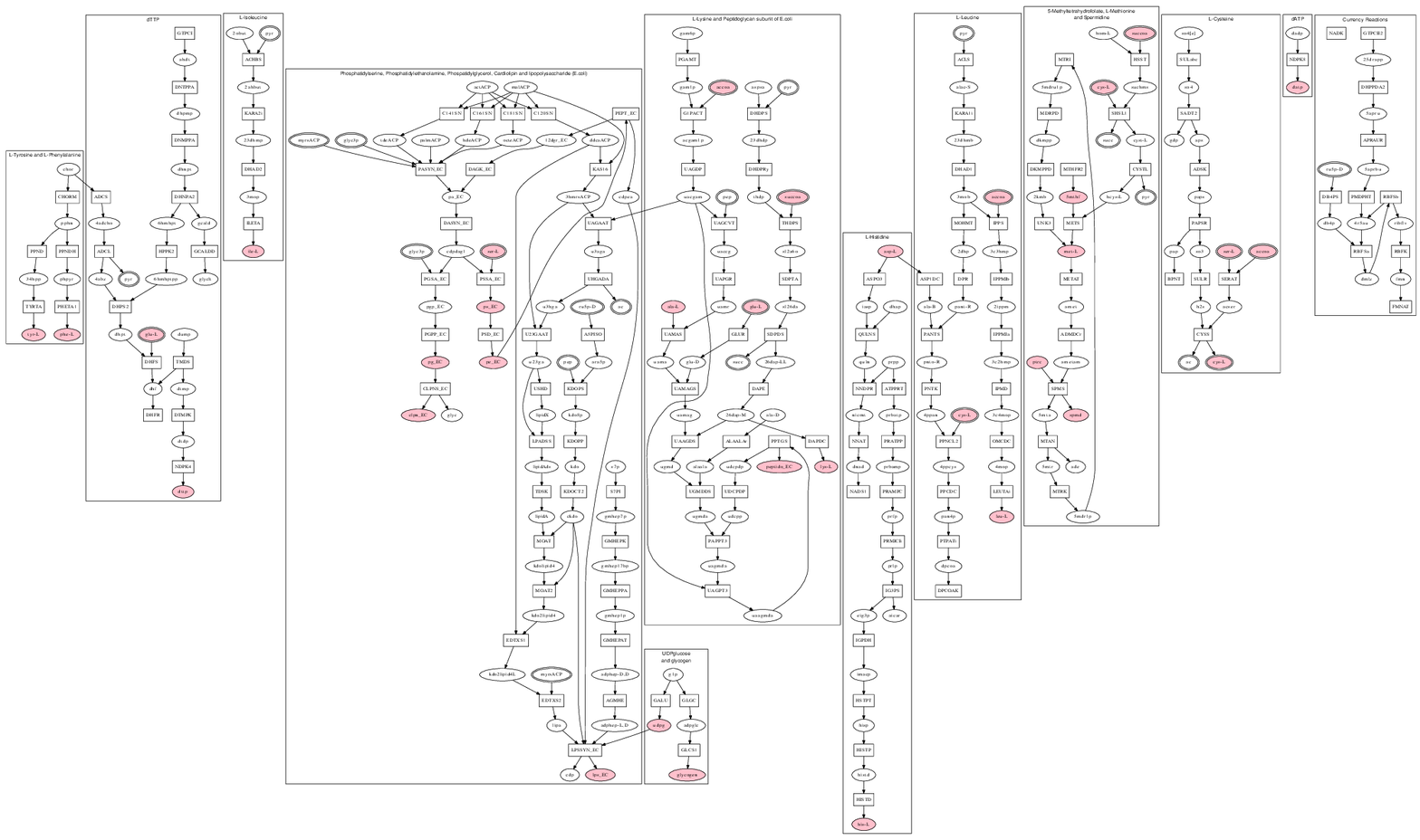}
\caption{\label{Fig5} {\bf Category IIa reactions in \emph{E.
coli}.} This figure shows the graph of 147 category IIa reactions in
\emph{E. coli} whose reaction fluxes are perfectly correlated across
minimal environments. Conventions are the same as in Figure 3. The
preponderance of biomass metabolites (pink ovals) in this figure
signifies that these reactions are at the output end of the
metabolic network. The reactions have been grouped together into
boxes based on common biosynthetic pathways. }
\end{figure*}

\begin{figure*}
\includegraphics[width=16cm]{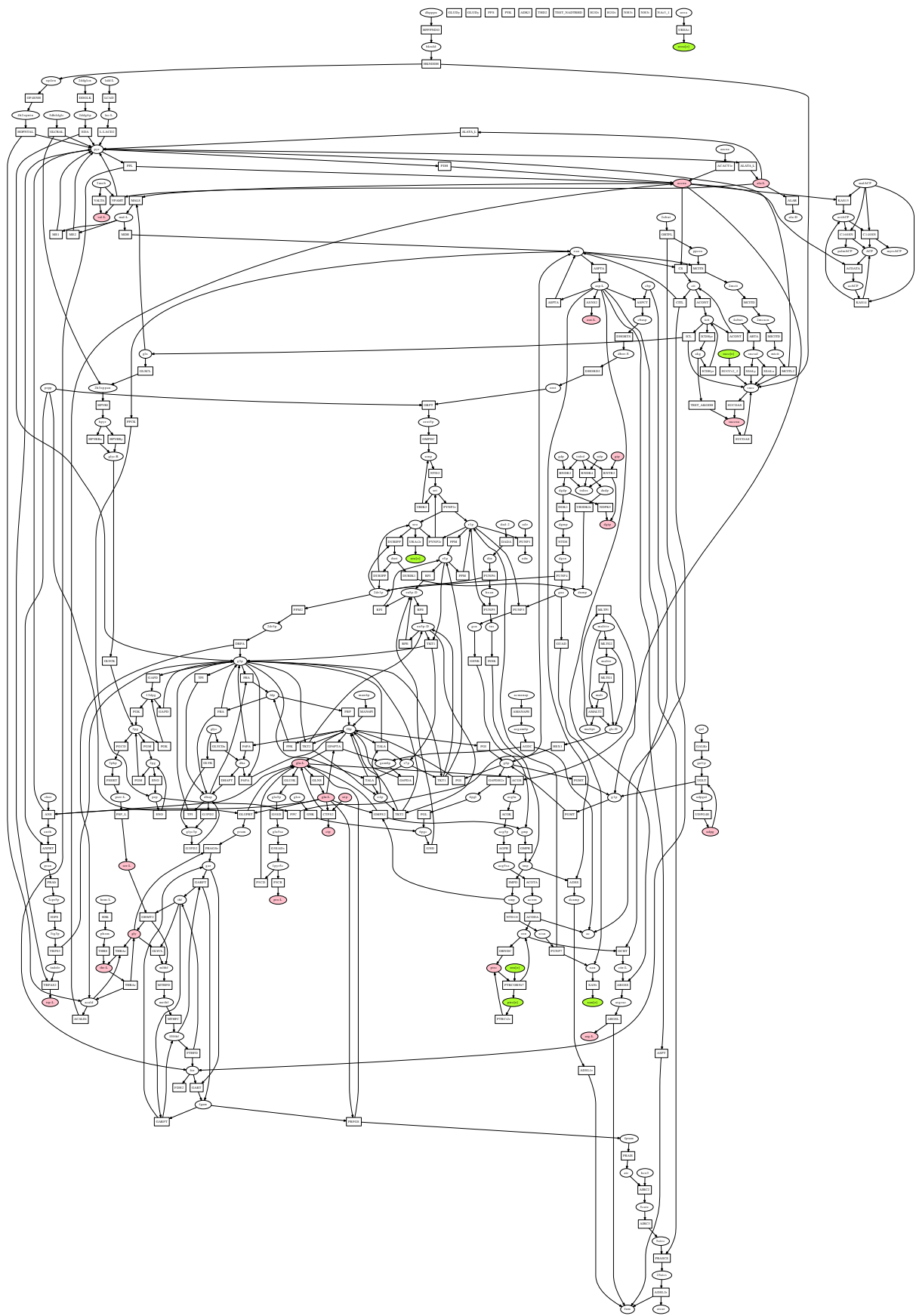}
\caption{\label{Fig6} {\bf Category III reactions in \emph{E.
coli}.} This figure shows the network of reactions that are active
in two or more minimal environments considered, but not in all the
environments. Conventions are the same as in Figure 3. Comparing
this graph of category III reactions with category I and IIa
reactions (cf. Figures 3 and 5), it is evident that category III
sub-network has a highly reticulate structure with many loops. }
\end{figure*}

\subsection{Category II: Output biosynthetic pathways} A key
biological function of the metabolic network is to convert nutrient
metabolites in the environment into biomass metabolites required for
growth and maintenance of the cell. The biomass metabolites, which
include all the amino acids, nucleotides, lipids and certain
cofactors, may be considered to be the output of the metabolic
network. Category II reactions are always-active and have a nonzero
flux for any feasible minimal environment. We found that the
category II sub-network has biosynthetic pathways for 30 out of the
49 biomass metabolites in \emph{E. coli}. These pathways are
typically the sole production pathways of those biomass metabolites
in \emph{E. coli} \cite{samal2006low}. Thus, this sub-network is at
the output end of the metabolism.

Of the 189 category II reactions in \emph{E. coli}, 147 reactions
belong to category IIa, whose fluxes are perfectly correlated across
the different minimal environments. Fig. 5 shows the graph of the
category IIa sub-network in \emph{E. coli}, which is the single
largest perfect cluster of reactions. The remaining 42 reactions in
category II constitute the category IIb, which are always active but
not perfectly correlated with category IIa reactions and with each
other. Thus, the fluxes of category IIb reactions vary in a more
complicated manner across minimal environments. Categories IIa and
IIb exist with similar properties in the metabolic networks of the
other two organisms (cf. Table \ref{stat}). In our previous work, we
have shown that most of the category II reactions are essential for
growth irrespective of the environment \cite{samal2006low}. The set
of category II reactions is a superset of reactions in the activity
core found earlier by Almaas {\it et al} \cite{almaas2005activity}
which are reactions always used across minimal as well as rich
environments.

\subsection{Category III: Intermediate pathways between input and
output} Fig. 6 shows the sub-network of category III reactions in
\emph{E. coli}, which are neither once-active nor always active; the
activity of these reactions depends on the availability of nutrients
in a more complicated manner. Category III reactions may be
considered to constitute the intermediate part of the network. By
comparing the structures of the three categories, it is evident that
category III has a highly reticulate and complex architecture
compared to category I and II. There is a functional reason for the
observed complexity in category III sub-network. The biomass
metabolites collectively contain several different types of chemical
structures (moieties), and the \emph{E. coli} metabolic network is
capable of producing these biomass metabolites from different
minimal environments, each containing a different (and single)
carbon source. A typical external carbon source has one or a few
moieties with different nutrients containing different subsets of
moieties. Category I reactions transport the carbon sources into the
cell and break it down into a small set of moieties. The function of
category III reactions is to start with a small set of moieties and
produce all the moieties required for biomass production. This
requires a complex set of internal transformations and the exact set
of transformations required depends on the nature of the input
moieties. Thus, the activity of category III transforming reactions
depends upon the biochemical nature of available nutrients in
different minimal environments. We find that category III contains
most of the reactions in central metabolism such as the citric acid
cycle. A similar architecture of category III sub-network was found
in the metabolic networks of the other two organisms as well. Some
of the biomass metabolites are produced in category III itself. For
the other biomass metabolites category III produces precursors which
are then taken up in the biosynthetic pathways of category II to
produce the biomass metabolites.

\section{Comparison of functional bow-tie decomposition with
graph-theoretic bow-tie decomposition}

Ma and Zeng \cite{ma2003connectivity,ma2004decomposition} have used
graph-theoretic measures to reveal a bow-tie architecture of
metabolic networks similar to that seen in World Wide Web (WWW)
\cite{broder2000graph}, wherein the network can be decomposed into
an in-component, out-component and a giant strong component. Given a
directed graph, a strong component is a maximal subgraph such that
for any pair of nodes $i$ and $j$ in the set there exists a directed
path from $i$ to $j$ and from $j$ to $i$ within the subgraph. In
general, a directed graph can have many strong components, and the
strong component with the largest number of nodes is designated as
the giant strong component (GSC). The associated in-component
consists of nodes which have access to GSC nodes via some directed
path, but cannot be reached from any GSC node via a directed path.
The out-component consists of nodes which can be reached from the
GSC nodes via some directed path, but lack access to any GSC node
via a directed path.

In this work, we have decomposed the metabolic network into three
categories using a simple algorithm based on activity patterns of
reactions across different minimal environments. Our categorization
reveals a functional bow-tie architecture wherein the input pathways
(category I reactions) fan into intermediate metabolism (category
III reactions) which forms the knot of a bow-tie and from where the
output pathways (category II reactions) for various biomass
components fan out.

In our functional bow-tie decomposition, the three categories I, II
and III of reactions discussed above broadly correspond to the
in-component, out-component and GSC, respectively, of the
graph-theoretic bow-tie decomposition by Ma and Zeng
\cite{ma2003connectivity,ma2004decomposition}. However, the
corresponding sets of reactions in the two decompositions differ in
detail. For example, we find that the end products of several (and
often long) chains of reactions in the category II sub-network are
re-cycled resulting in feedback loops. Such feedback loops in
category II sub-network presumably minimize wastage and could be
instrumental in producing the biomass metabolites in the desired
ratios. An example of such a feedback loop in category II
sub-network is the one involving metabolite 5mdr1p (which can be seen in the electronic version of Fig. 5 upon zooming). The
biosynthetic pathways involved in such feedback loops appropriately belong to the
output part of metabolism because they connect the precursor
metabolites to the outputs. However, the graph-theoretic bow-tie
decomposition would classify such category II reactions in feedback
loops into the GSC. Thus, our functional bow-tie decomposition based
on fluxes of reactions across different environments gives better a
insight and is biochemically more realistic. The picture of the
metabolic network our decomposition reveals is similar in spirit to
the one envisioned by Csete and Doyle \cite{csete2004bow}. Further,
it is important to note that our flux-based categorization does not
involve the a priori exclusion of high degree currency metabolites
as was needed in the graph-theoretic bow-tie decomposition of the
metabolic network \cite{ma2003connectivity,ma2004decomposition}.

\section{Conclusions}
In this paper, we have performed flux balance analysis (FBA) for the
metabolic networks of three microorganisms: \emph{E. coli}, \emph{S.
cerevisiae} and \emph{S. aureus} to obtain fluxes of reactions in
the network under diverse environmental conditions. We have followed
a purely algorithmic approach leveraging on the predicted fluxes of
reactions across different minimal environments to decompose the
metabolic network into functionally relevant sub-networks. We find
that the activity of a reaction given by the number of minimal
environments for which it has a nonzero flux is an important
indicator of the functional role of a reaction. We have classified
the reactions into three functional categories based on their
activity. Category I contains once-active reactions which are used
in only one minimal environment. Most reactions belonging to the
category I sub-network are uptake pathways for external nutrients in
feasible minimal environments, and the primary function of these
reactions is to catabolize external nutrients into simpler
metabolites which can be further processed by intermediary
metabolism. Category II contains always active reactions which are
used in all minimal environments. The category II sub-network is
critical for the survival of the organism and accounts for the
majority of the biosynthetic pathways for the production of the
biomass metabolites at the output end of metabolic network. Category
III contains reactions which are used in an intermediate number of
minimal environments, and is responsible for generating the
`precursor' molecules that are eventually converted into biomass
metabolites by Category II reactions. We find that while category I
and II sub-networks are dominated by simple linear pathways, the
structure of the category III sub-network is highly reticulate. In
summary, our decomposition method for large-scale metabolic networks
based on activity of reactions captures the proposed functional
bow-tie organization by Csete and Doyle: the input pathways
(category I reactions) for nutrients in the environment fan into
intermediate metabolism (category III reactions) which forms the
knot of bow-tie from where the output biosynthetic pathways
(category II reactions) for biomass components fan out. Our results
are valid for metabolic networks of three phylogenetically different
organisms (two distinct prokaryotes and a eukaryote), which suggests
that the observed functional bow-tie organization could be quite
common in living systems.


\appendix

\section{Robustness of categorization of reactions to alternate
optimal solutions}

In this work, flux balance analysis (FBA) was used to obtain a
particular flux vector ${\bf v}$ or optimal solution that maximizes
the objective function taken as the growth rate in a given minimal
environment. However, for large-scale metabolic networks, there
exist multiple flux vectors ${\bf v}$ or alternate optimal solutions
that maximize growth in a given minimal environment, i.e., there are
many flux vectors ${\bf v}$  with exactly the same value of the
objective function but use different alternate pathways in the
network
\cite{lee2000recursive,mahadevan2003effects,reed2004genome,samal2008conservation}.
FBA finds one of many possible alternate optima for a given minimal
environment that maximizes growth. In the main text, we have used a
single optimal flux vector ${\bf v}$  for each of the $M$ feasible
minimal environments to determine the activity of a reaction and the
set of active reactions in the metabolic network of an organism.
Since, in principle, the activity of a reaction can change depending
on the particular flux vector considered, we study the robustness of
our categorization of reactions to the presence of alternate optima.

Flux variability analysis (FVA) \cite{mahadevan2003effects} can be
used to determine the set of reactions whose fluxes vary across
alternate optima for a given minimal environment. Specifically, FVA
determines the maximum and minimum flux value that each reaction can
take across alternate optima for a given minimal environment. FVA
involves the following steps:
\begin{itemize}
\item[(a)] Determine using FBA the maximum value of the objective
function $Z$ or growth rate $v_{biomass}^\alpha$ in a given minimal
environment $\alpha$. \item[(b)] Fix the flux of the biomass
reaction equal to $v_{biomass}^\alpha$. \item[(c)] Change the
objective function $Z$ to be the flux of a reaction $j$. \item[(d)]
Using linear programming determine the maximum flux value
$v_{j,max}^\alpha$ of reaction $j$ in the minimal environment
$\alpha$, constraining the biomass reaction to have a flux equal to
$v_{biomass}^\alpha$.
\item[(e)] Using linear programming determine the minimum flux value
$v_{j,min}^\alpha$ of reaction $j$ in the minimal environment
$\alpha$, constraining the biomass reaction to have a flux equal to
$v_{biomass}^\alpha$.
\item[(f)] The range $v_{j,min}^\alpha$ to $v_{j,max}^\alpha$ gives
the variability of flux of reaction $j$ across different alternate
optima. \item[(g)] The above steps c, d, e and f can be repeated for
every reaction $j$ in the metabolic network to determine the flux
variability of each reaction across alternate optima for a given
minimal environment $\alpha$.
\end{itemize}

We have used FVA to determine $v_{j,max}^\alpha$ and
$v_{j,min}^\alpha$ for each reaction $j$ and for each feasible
minimal environment $\alpha$ in the \emph{E. coli} metabolic
network. A reaction $j$ is designated as \emph{blocked} if
$v_{j,max}^\alpha$=0 for all $M$ feasible minimal environments
\cite{schuster1991detecting,burgard2004flux}. We found 329 blocked
reactions in the \emph{E. coli} metabolic network. The remaining 838
reactions, for which $v_{j,max}^\alpha$$>$0 for at least some
environment $\alpha$ are designated as \emph{potentially active}
reactions. This set includes the 585 active reactions considered in
the main text. We define a reaction $j$ as \emph{essential} for a
given minimal environment $\alpha$ if $v_{j,min}^\alpha$$>$0. 484
reactions were found to be essential for some $\alpha$ in the
\emph{E. coli} metabolic network which are a subset of the 585
active reactions considered in the main text. We now classify these
484 reactions into the following three categories:
\begin{itemize}
\item[(a)] Essential category I: Reactions which satisfy $v_{j,min}^\alpha$$>$0 for exactly
one minimal environment. We found 162 reactions in the \emph{E.
coli} metabolic network to be in Essential category I. Of these, 153
reactions belong to category I of the main text.
\item[(b)] Essential category II: Reactions which satisfy $v_{j,min}^\alpha$$>$0 for all $M$ minimal environments. We found
171 reactions in the \emph{E. coli} metabolic network to be in
Essential category II. All of these belong to category II of the
main text. \item[(c)] Essential category III: Reactions which
satisfy $v_{j,min}^\alpha$$>$0 for $m$ minimal environments where
1$<$$m$$<$$M$. We found 151 reactions in the \emph{E. coli}
metabolic network to be in Essential category III. Of these, 145
belong to category III of the main text.
\end{itemize}
Thus we find that the classification discussed in the main text
which uses a particular flux vector correctly predicts the essential
category I, II or III of 469 out of the 484 essential reactions.

\section{Relation between standard deviation $\sigma$ and mean flux $\langle v \rangle$ for
category I and category IIa reactions}

In Fig. 2, we plot the standard deviation $\sigma$ versus mean flux
$\langle v \rangle$ for active reactions in a metabolic network
across its $M$ feasible minimal environments. Here, we derive the
relation between mean flux $\langle v \rangle$ and standard
deviation $\sigma$ for reactions in category I and category IIa
shown as upper and lower lines, respectively, in Fig. 2.

\subsection{Category I reactions} In a given organism any reaction
belonging to category I has activity $m$=1, and is active for a
single environment (say $\alpha_0$). The mean flux $\langle v_j
\rangle$ of a category I reaction $j$ across $M$ feasible
environments is given by:
\begin{eqnarray}
\label{mean1} \langle v_j \rangle &=& \frac{1}{M}
\sum_{\alpha=1}^{M} v_j^{\alpha} \nonumber \\ &=&
\frac{v_j^{\alpha_0}}{M} ,
\end{eqnarray}
where $v_j^\alpha$ is the flux of reaction $j$ in the environment
$\alpha$ $(\alpha=1,2,\ldots,M)$. $v_j^{\alpha_0}$ is the flux of
reaction $j$  in the only feasible minimal environment $\alpha_0$
where the reaction has nonzero value and in all other feasible
minimal environments the flux of reaction $j$ is 0.

Thus, the standard deviation $\sigma_j$ for a category I reaction
$j$ is given by:
\begin{eqnarray}
\sigma_j &=& \sqrt{\frac{1}{M} \sum_{\alpha=1}^{M} (v_j^\alpha -
\langle v_j \rangle)^2} \nonumber \\ &=& \sqrt{\frac{1}{M}[(M-1)
\langle v_j \rangle^2 + (v_j^{\alpha_0} - \langle v_j \rangle)^2] }
\nonumber
\\ &=& \sqrt{M-1}\langle v_j \rangle ,
\end{eqnarray}
where we have used the result in Eq. \ref{mean1}.

\subsection{Category IIa reactions}
For two reactions $j$ and $k$ in category IIa, their flux
$v_j^\alpha$ and $v_k^\alpha$ in a given environment $\alpha$ are
perfectly correlated to each other. The fluxes of category IIa
reactions are proportional to each other having the same
proportionality constant for all minimal environments. For a minimal
environment $\alpha$, we can write the flux of category IIa reaction
$j$ as:
\begin{equation}
v_j^\alpha=c^\alpha v_j^0 ,
\end{equation}
where $c^\alpha$  is a constant for the minimal environment $\alpha$
and $v_j^0$ is some number. For any two reactions $j$ and $k$ in
category IIa with fluxes correlated across minimal environments, we
have:
\begin{eqnarray}
\frac{v_j^\alpha}{v_k^\alpha} &=& \frac{c^\alpha v_j^0}{c^\alpha
v_k^0} \nonumber \\ &=&
\frac{v_j^{\alpha^\prime}}{v_k^{\alpha^\prime}} ,
\end{eqnarray}
where $\alpha$ and $\alpha^\prime$ are two different feasible
minimal environments for the organism.

The mean flux of reaction $j$ is:
\begin{eqnarray}
\label{mean2a} \langle v_j \rangle &=& \frac{1}{M}
\sum_{\alpha=1}^{M} v_j^\alpha \nonumber \\ &=& v_j^0 \frac{1}{M}
\sum_{\alpha=1}^{M} c^\alpha \nonumber \\ &=& v_j^0 \langle c
\rangle ,
\end{eqnarray}
where $\langle c \rangle$ is the mean of $c^\alpha$ across the set
of feasible minimal environments.

The standard deviation $\sigma_j$ for category IIa reaction $j$ is
given by:
\begin{eqnarray}
\sigma_j &=& \sqrt{\frac{1}{M} \sum_{\alpha=1}^{M} (v_j^\alpha -
\langle v_j \rangle)^2} \nonumber \\ &=& v_j^0 \sqrt{\frac{1}{M}
\sum_{\alpha=1}^{M} (c^\alpha - \langle c \rangle)^2} \nonumber \\
&=& v_j^0 \sigma_c \nonumber \\ &=& \frac{\sigma_c \langle v_j
\rangle}{\langle c \rangle} \nonumber \\ &=& b \langle v_j \rangle ,
\end{eqnarray}
where we have used the result in Eq. \ref{mean2a}.

\begin{acknowledgments}
SS and VG acknowledge support from University Grants Commission
(UGC), AS from Council for Scientific and Industrial Research
(CSIR), and SJ from Department of Biotechnology (DBT), India.
\end{acknowledgments}


%

\end{document}